\documentstyle[prl,aps,preprint]{revtex}
\def\thru#1{\mathrel{\mathop{#1\!\!\!/}}}

\begin{document}
\tighten

\title{$Q^2$-DEPENDENCE OF THE NUCLEON'S \\
$G_1$ STRUCTURE FUNCTION SUM RULE\thanks
{This work is supported in part by funds provided by the U.S.
Department of Energy (D.O.E.) under contract \#DE-AC02-76ER03069.}}

\author{Xiangdong Ji and Peter Unrau}

\address{Center for Theoretical Physics \\
Laboratory for Nuclear Science \\
and Department of Physics \\
Massachusetts Institute of Technology \\
Cambridge, Massachusetts 02139 \\
{~}}

\date{hep-ph/9308263 ~~~ MIT-CTP-2232 ~~~ %
Submitted to {\it Phys. Rev. Lett.} ~~~ August 1993}

\maketitle
\setcounter{page}{0}
\thispagestyle{empty}

\begin{abstract}
We study the $Q^2$ variation of the first moment of the nucleon's
spin-dependent structure function $G_1$. As $Q^2 \rightarrow 0$ the moment is
determined by the low energy theorem for Compton scattering. In
the deep-inelastic region the moment is calculated using twist expansion to
order $1/Q^2$. Based on these limits, we construct a formula which smoothly
interpolates between the two regions.
\end{abstract}


\narrowtext
Recently, polarized deep-inelastic scattering has
proven to be an excellent tool for studying the spin
structure physics of the nucleon \cite{EMC,E142,SMC}.
Supplemented with the operator product expansion
analysis in Quantum Chromodynamics (QCD),
experimental data at high-energy provides a direct
measurement of the matrix elements
of spin-dependent operators in the nucleon.
A much discussed example in the current
literature is the axial charge, or the forward matrix element of
axial current, whose measurement by the EMC collaboration
casts doubt on our traditional understanding
of the proton's spin structure \cite{EMC}.

A closely related question is can one learn
anything about the nucleon's spin structure from
electro-production experiments away from
the deep-inelastic limit? In particular, what insight
do the spin structure functions $G_1$ and $G_2$ provide
at low and moderate $Q^2$? Not long ago, Anselmino {\it et al.} \cite{AIL}
pointed out that at the real photon point
($Q^2= 0$), the first moment of $G_1$
(called the sum rule in the following text) is
related, via the celebrated Drell-Hearn-Gerasimov (DHG)
sum rule \cite{DHG}, to the anomalous magnetic moment of
the nucleon, and thus the
physics of the $G_1$ structure function again appears simple
in the $Q^2 \rightarrow 0$ limit. Together with
knowledge from the deep-inelastic limit, the authors in Ref.~\cite{AIL}
constructed a model for the sum rule at all $Q^2$.
This has motivated a number of proposals
to measure $G_1$ and $G_2$ at low energy \cite{AHR,BUR}.

In Ref.~\cite{JI1}, one of us pointed out that the
analysis made in Ref.~\cite{AIL} excluded
the nucleon's elastic contribution to the moment,
which in the $Q^2\rightarrow 0$ limit dominates the entire
inelastic contribution calculated from the
DHG sum rule. He argued that the moment has to
include this contribution if it
is to be analyzed in twist expansion in
the deep-inelastic limit and its experimental
measurement is to be used to extract the matrix elements
of higher-twist operators.

In this Letter we study the
$Q^2$ variation of the sum rule by exploring the physics
of the $Q^2\rightarrow 0$ and $Q^2 \rightarrow \infty$
limits. In the first limit, we rely on the low energy
result derived in Ref.~\cite{JI1}, to calculate the exact
value and the first derivative of the sum rule at $Q^2=0$.
In the second limit,
we use a twist expansion appropriate for the
deep inelastic region, focusing on the $1/Q^2$
correction term. The matrix elements of higher
twist operators are related to
moments of the quark distributions functions
$g_{if}(x) (i=1, T, 3$ and $f=u,d,s...)$
through a novel use of the QCD equations of motion,
which are in turn evaluated in the MIT bag model.
As an application, we discuss the correction of
higher twists to the Bjorken sum rule in the deep
inelastic limit. Having obtained
analytical results valid for the
low and high ranges of $Q^{2}$, we construct a simple
parameterization to smoothly interpolate
both limits, which should be checked experimentally.

To begin we consider the following fixed-mass sum rule,
\begin{equation}
\Gamma(Q^2) = {Q^2\over 2M^2}
\int^{\infty}_{Q^2/2} G_1 (\nu, Q^2) {d\nu \over \nu},
\label{sum1}
\end{equation}
where $G_1(\nu, Q^2)$ is one of the nucleon's spin dependent
structure functions in the nucleon tensor,
\begin{equation}
W_{\mu\nu}^A = -i\epsilon_{\mu\nu\alpha\beta}
q^{\alpha}\Big[S^{\beta} {G_1\over M^2} + {G_2 \over M^4}
(\nu S^{\beta} - P^{\beta} (S\cdot q)) \Big].
\label{wmunu}
\end{equation}
The lower integration limit in Eq. (\ref{sum1}) implies the
elastic contribution to $G_1$ is also included.
Here $P$ and $S$ are the nucleon's momentum and polarization,
$q$ is the virtual photon momentum, $M$ is the nucleon mass, $\nu = P\cdot q$
and $Q^2=-q^2$ ($\epsilon^{0123}=1)$. In deep-inelastic limit, one defines
scaling functions $g_1(x, Q^2) = \nu/M^2 G_1$
and $g_2(x, Q^2) = (\nu/M^2)^2 G_2$.
The sum rule then becomes,
\begin{equation}
\Gamma(Q^2) = \int^1_0 g_1(x, Q^2) dx,
\label{sum2}
\end{equation}
which is just the first moment of the scaling function.

Let us first consider the small $Q^2$ behavior of the sum rule.
Introduce a spin-dependent virtual-photon Compton amplitude
$S_1(\nu, Q^2)$ whose imaginary part is proportional to $G_1$,
and write down the unsubtracted dispersion relation,
\begin{equation}
       S_1(\nu, Q^2) = 4\int^{\infty}_{Q^2/2}
               {\nu'd\nu' \over \nu'^2-\nu^2} G_1(\nu', Q^2).
\label{disp}
\end{equation}
Through this, we relate the sum rule to the Compton amplitude
at $\nu=0$,
\begin{equation}
           \Gamma (Q^2) = {Q^2\over 8M^2} S_1(0, Q^2).
\label{rela}
\end{equation}
At small $\nu$ and $Q^2$, the dominant contribution to
$S_1$ comes from the nucleon pole diagrams \cite{JI1},
\begin{equation}
    S_1^{\rm pole}(\nu, Q^2) = -2M^2F_1(F_1+F_2)
            [{1\over 2\nu-Q^2} - {1\over 2\nu+Q^2}]
        - F_2^2,
\label{pole}
\end{equation}
where $F_1$ and $F_2$ are the Dirac and
Pauli form factors of the nucleon.
 From this, we obtain for $Q^2\rightarrow 0$,
\begin{equation}
     \Gamma(Q^2) = {1\over 2}F_1(F_1 + F_2) - {1\over 8M^2}F_2^2Q^2.
\label{lowt}
\end{equation}
This result can be shown to be accurate
up to the order of $Q^2$ in the small
$Q^2$ region by {\it explicitly}
evaluating Eq. (\ref{sum1}): The elastic
contribution to $G_1$ is proportional to $\delta
(2\nu-Q^2)$, producing the first term in Eq. (\ref{lowt});
the integral over inelastic contributions
is just the DHG sum rule in the limit of
$Q^2\rightarrow 0$ and the second term in
Eq. (\ref{lowt}) reproduces this in the same limit.

The elastic contribution vanishes
identically at $Q^2=0$ because of energy-momentum
conservation, and the DHG sum rule indicates
$\Gamma(0)=0$. Thus, due to the elastic contribution,
$\Gamma(Q^2)$ is
non-analytic around $Q^2=0$, i.e.,
\begin{equation}
        \Gamma(Q^2=0) \ne  \Gamma(Q^2\rightarrow 0) .
\label{nona}
\end{equation}
To remedy this, one can take two approaches:
The first approach
subtracts away the elastic
contribution from the sum rule for $Q^2 \ne 0$.
The new sum,  $\bar \Gamma (Q^2) =
\Gamma(Q^2) - 1/2F_1(F_1+F_2)$), is a
smooth extension of the DHG sum rule to
virtual-photon scattering. The approach
we take in this paper is to redefine
$\Gamma$ at $Q^2=0$,
\begin{equation}
        \Gamma(Q^2=0) \equiv  \Gamma(Q^2\rightarrow 0).
\label{exte}
\end{equation}
This approach ensures that the sum rule at low $Q^2$
can be treated with the twist
expansion that we will discuss below.
The expansion is for moments of
the $g_1$ structure function which include the
integration limit $x=1$, where the elastic
contribution resides.

Since Eq.~(\ref{lowt}) is accurate up to the order of
$Q^2$, we can determine $\Gamma(Q^2)$ and
its first derivative in $Q^2\rightarrow 0$ limit,
\begin{eqnarray}
\Gamma^p(0) &=& 1.396, \nonumber \\
\Gamma^n(0) &=& 0, \nonumber \\
{d\Gamma^p(Q^2) \over dQ^2} |_{Q^2 =0} &=&  -8.631 {\rm GeV}^{-2}, \nonumber \\
{d\Gamma^n(Q^2) \over dQ^2} |_{Q^2 =0} &=&  -0.479 {\rm GeV}^{-2},
\label{gamd}
\end{eqnarray}
where $p,n$ refer to proton or neutron and
the squares of the proton and neutron charge radii
$\langle r^2_p\rangle_{\rm c.r.}= (0.862~{\rm fm})^2 $ and
$\langle r^2_n\rangle_{\rm c.r.}= -(0.342~{\rm fm})^2$ have been used.
The initial slope of $\Gamma^p(Q^2)$ is
primarily determined by the elastic contribution as
the inelastic contribution, ${-\kappa^2_p/ 8M^2} = -0.455
{\rm GeV}^{-2} $, is only about 5\% of the total.
Therefore, one expects that for small $Q^2$,
$\Gamma^p(Q^2)$ is mainly given by the elastic contribution.
In contrast, due to a numerical coincidence,
the elastic part of $\Gamma^n(Q^2)$
is negligible compared with the inelastic part.

In the limit of large $Q^2$ ($Q^2>\!\!>\Lambda_{\rm QCD}^2$),
$\Gamma(Q^2)$ can be calculated in terms of the twist expansion,
\begin{equation}
\Gamma(Q^2) = \sum_{\tau=2,4...}  {\mu_\tau(Q^2)\over (Q^2)^{\tau-2 \over 2}},
\label{twie}
\end{equation}
where $\mu_\tau (Q^2)$ are matrix elements of quark-gluon
operators which scale like
$\Lambda_{\rm QCD}^{\tau-2}$. The $Q^2$-dependence in
$\mu_\tau$ are logarithmic and can be calculated
in perturbative QCD. If the nucleon mass were zero,
$\mu_\tau(Q^2)$ would contain only twist-$\tau$ operators.
The effect of the nucleon mass is to induce contributions to
$\mu_\tau(Q^2)$ from lower twist operators, as we shall illustrate
below.

The leading term in Eq.~(\ref{twie}) is well-known,
\begin{equation}
\mu_2 = {1\over 2} \sum_{f=u,d,s...} e_f^2 a_{0f},
\label{eq:8}
\end{equation}
where the summation covers quarks of all flavors $f$ and
$a_{0f}$ is the axial charge defined by the
matrix element of axial current
$A^\mu_f = \bar \psi_f \gamma^\mu\gamma_5 \psi_f$:
$\langle PS|A^\mu_f|PS\rangle  = 2a_{0f} S^\mu$. The
QCD radiative corrections have been calculated to
the first order in $\alpha_s(Q^2)$ ~\cite{KMSU}
for the singlet contribution ($a^{S}_0 =
2(a_{0u}+a_{0d}+a_{0s})/9$) and to the third order
{}~\cite{LTV} for the non-singlet contribution ($a^{NS}_0
= (2a_{0u}-a_{0d}-a_{0s})/9$).
The proton-neutron difference of
the moment defines the Bjorken sum rule,
\begin{equation}
\mu^p_2(Q^2) - \mu^n_2(Q^2) =
{g_A\over 6} \left[1-{\alpha_s(Q^2)\over \pi}-3.58
  \left({\alpha_s(Q^2)\over \pi}\right)^2 -
20.2\left({\alpha_s(Q^2)\over \pi}\right)^3+...\right],
\label{bj}
\end{equation}
where $g_A=a_{0u}-a_{0d} = 1.257$ is the neutron decay constant.

The $1/Q^2$ power corrections to $\Gamma$
were first studied by Shuryak and Vainshtein (SV) \cite{SV}.
Using the collinear expansion technique \cite{EFP},
one of us has calculated in Ref. \cite{JI2} the entire
$1/Q^2$ corrections to the $g_1$ scaling function in
terms of a few multi-parton distribution
functions. Specializing to the
first moment, we find
\begin{eqnarray}
\mu_4 &=& {1\over 9} \sum_f e_f^2\Big[a_{2f}
+ d_{2f} - 4f_{2f} \Big]M^2 \nonumber\\
&=& (A+D+F)M^2.
\label{resl}
\end{eqnarray}
where $A={1\over 9} \sum_f e_f^2 a_{2f}$
comes from the twist-two contribution and
$a_{2f}$ is defined as,
\begin{equation}
\langle PS|\bar \psi_f\gamma^{(\sigma}\gamma_5
  iD^{\mu_1}iD^{\mu_2)}\psi_f|PS\rangle
 = 2a_{2f} S^{(\sigma}P^{\mu_1}P^{\mu_2)},
\label{twist2}
\end{equation}
with $(\cdot\cdot\cdot)$ denotes symmetrizing the indices
and subtracting the trace;
$D={1\over 9} \sum_f e_f^2 d_{2f}$ comes from the twist-three
contribution and $d_{2f}$ is defined as,
\begin{equation}
\langle PS|g\bar \psi \tilde F^{\sigma(\mu_1}\gamma^{\mu_2)} \psi |PS\rangle
   = 2d_{2f}S^{[\sigma} P^{(\mu_1]} P^{\mu_2)},
\label{twist3}
\end{equation}
with $[\cdot\cdot\cdot]$ denotes anti-symmetrizing the
indices and $\tilde F^{\sigma\mu_1} = 1/2
\epsilon^{\sigma\mu_1\alpha\beta}F_{\alpha\beta}$
is the dual of the gluon field tensor;
$F= -{4\over 9} \sum_f e_f^2 f_{2f}$ comes from the
twist-four contribution and $f_{2f}$ is
defined as,
\begin{equation}
\langle PS|g\bar \psi_f \tilde F^{\mu\nu}
\gamma_\nu \psi_f|PS\rangle  = 2 f_{2f} M^2 S^\mu.
\label{twist4}
\end{equation}
We note that the result quoted for $\mu_4$
in Ref.~\cite{SV} is $(2A+2D+F)M^2.$

To study the QCD radiative corrections to $\mu_4$, one has
to consider operator-mixing from
gluon operators, the anomalous dimensions of
which are not currently
available and their matrix elements
are difficult to estimate. Therefore, in the
following discussion, we neglect
entirely the scale dependence of
$\mu_4$.

The higher twist operators in Eqs.~(\ref{twist3})
and~(\ref{twist4})
depend explicitly on gauge fields. To calculate their
matrix elements we need a wave function of
the nucleon containing gluon components. However, for
special types of higher twist operators such as the present case,
we can eliminate the gluons in terms of
the ``bad'' components of quark fields using
the QCD equations of motion \cite{JI2}.
Then the higher twist matrix elements
can be related to moments of parton distributions
with no explicit gluon fields. Indeed, by defining in the
light-cone gauge ($A\cdot n = 0$),
\begin{equation}
g_{(1,T,3)f}(x) = {1\over 2}
  \int {d\lambda \over 2\pi} e^{i\lambda x}
  \langle PS|\bar \psi_f \thru
 Q_{(1,T,3)} \gamma_5 \psi_f(\lambda n)|PS \rangle,
\end{equation}
where $Q_1 = n, Q_T = -S_T/M, Q_3 = -2p/M^2$
and $n$ and $p$
are two null vectors ($n^2=p^2=0$ and $p\cdot n = 1$),
we find,
\begin{eqnarray}
d_{2f} &= &{1\over 2} \int x^2(3g_{2f}(x) + 2g_{1f}(x)) dx, \nonumber \\
   f_{2f}& = &{1\over 2}\int x^2 (7g_{1f}(x)
   + 12g_{2f}(x) - 9g_{3f}(x)) dx,
\label{conv}
\end{eqnarray}
where $g_{2f} = g_{Tf}-g_{1f}$.
These relations are exact in QCD.

We choose to estimate the $1/Q^2$ corrections to
the sum rule in the simplest version of the
MIT bag model, in which the bag boundary
simulates gluon confinement \cite{jaffe,jaffe2}.
Using $g_{if}(x)$ $(i=1,T,3)$
calculated in this model,
we obtain for the proton, $A^p= 0.0065$,
$D^p=0.0092 $, and $F^p= 0.0155$.
Inserting them into Eq.~(\ref{resl}), we have,
\begin{equation}
\mu_4^p({\rm Bag}) =  0.031 M^2.
\label{bag1}
\end{equation}
Compared with the size of $\mu_2^p=0.126 \pm 0.025$ from the
EMC data or 0.175 from the Ellis-Jaffe sum rule,
the bag $1/Q^2$ power correction is about 10\% at $Q^2=2$ GeV$^2$
and about 2\% at $Q^2=10$ GeV$^2$.
Assuming there are no abnormal twist-six or higher
contributions, we conclude that most $Q^2$
variations of the proton sum rule
occur below 1 GeV$^2$. For the neutron, the bag model
predicts, $A^n=D^n=F^n=0$, and the $1/Q^2$ correction
vanishes:
\begin{equation}
\mu_4^n({\rm Bag}) = 0.
\label {bag2}
\end{equation}
This follows from the SU(6) structure of
the bag wave function, which also predicts $\mu_2^n=0$.

The higher twist matrix elements have also been calculated
using the QCD sum rule (QSR) technique
by Balitsky {\it et al} \cite{BBK}.
Their most recent result in terms of our notation
is
\begin{eqnarray}
      \mu_4^p({\rm QSR})& = & -(0.023\pm 0.015)M^2, \nonumber\\
      \mu_4^n({\rm QSR})& = & -(0.006\pm 0.004)M^2.
\label{qsr}
\end{eqnarray}
Thus, the QCD sum rule calculation gives a power correction
the same size as the bag calculation with the correction
for the neutron being significantly smaller than for
the proton. However, the sign of the correction differs from
the bag result in Eq. (\ref{bag1}).
This difference has a large effect on the Bjorken sum rule
at small $Q^2$.

The Bjorken sum rule has recently been extracted
from the data on the proton \cite{EMC} and neutron \cite {E142}
$g_1$ structure functions,
\begin{eqnarray}
      \int^1_0 g_1^{p-n}(x, {\rm 2 ~GeV^2}) dx &= & 0.146 \pm 0.021
\cite{E142}, \nonumber\\
                              & = & 0.152 \pm 0.025 \cite{EK}.
\label{data}
\end{eqnarray}
In QCD, the Bjorken sum rule
at low $Q^2$ is contaminated by higher twist corrections
discussed above. If the QSR result is used for the correction,
one obtains a theoretical prediction at the same $Q^2$,
\begin{equation}
      \int^1_0 g_1^{p-n}(x, {\rm 2 ~GeV^2}) dx = 0.160 \cite{EK},
\label{qsr1}
\end{equation}
On the other hand, the bag result produces,
\begin{equation}
      \int^1_0 g_1^{p-n}(x, {\rm 2 ~ GeV^2}) dx = 0.182.
\end{equation}
While Eq. (\ref{qsr1}) gives a corrected
Bjorken sum rule within the experiment errors,
the bag calculation disagrees with the
extraction of the sum rule by the
E142 collaboration by 1.7$\sigma$
and with the extraction by Ellis-Karliner by 1.2$\sigma$.
In our opinion, a deviation
from the Bjorken sum rule means either a
measurement of higher twist matrix elements,
or that the data are inconsistent.

 From the high and low $Q^2$ knowledge of the sum rule, we
propose a model for $\Gamma^p(Q^2)$ in the entire $Q^2$ region,
\begin{equation}
\Gamma^p(Q^2) = {1\over 2}F_1(F_1+F_2)(1-\lambda_1 {Q^2\over M^2})
            + \lambda_2 {1 + \lambda_3 M^2/Q^2 \over
            1 + \lambda_4 M^4/Q^4},
\label{eq:31}
\end{equation}
where the first term is the elastic contribution
with its derivative modified by the $Q^2$ term.
The second term is basically $a + b/Q^2$ and the denominator
serves to suppress the contribution at small $Q^2$.
 From the EMC data and the various constraints derived above, we
determine all $\lambda_i$ except $\lambda_4$, which
controls the size of the twist-six contribution. The solid
and upper-dashed curves shown in Fig. 1 are
our parameterization with the bag and QSR higher
twist matrix elements, respectively.
[We choose $\lambda_4 = 0.3$, which gives a $\mu_6 \sim - 0.03$.]
The dotted curve represents the result of the twist expansion
to order $1/Q^2$ and the dot-dashed curve represents
the elastic contribution. As can be seen from the
figure, the different choices for higher twist matrix elements
result in about 15\% difference in $\Gamma$ in the $Q^2 = 0.5$ to
1.0 GeV$^2$ region. A similar interpolation is made for the
neutron, and the result is shown as the lower-dashed curve.

Thus it appears that the $Q^2$ variation of the $\Gamma(Q^2)$ sum rule
is quite simple. Nevertheless, its experimental measurement
is interesting, particularly around $Q^2=0.5$ GeV$^2$.
If we know $\Gamma(Q^2)$ in an extended $Q^2$ region, we
can fit data with a parameterization similar to the one used in
Eq.~(\ref{eq:31}). Then by expanding in a $1/Q^2$ power series, we
can extract the higher-twist matrix elements, such as
$f_{2f}$, which shall provide valuable insight into the
spin structure of the nucleon.

We thank Bob Jaffe for checking Eq.~(\ref{resl}) and
I. Balitsky and A. Vainshtein for several useful discussions.
PU acknowledges the support of a NSERC scholarship from the
Canadian government.


\end{document}